\documentclass[a4paper,11pt]{article}
\usepackage{pos}

\newcommand{\be}{\begin{equation}}
\newcommand{\ee}{\end{equation}}
\newcommand{\bea}{\begin{eqnarray}}
\newcommand{\eea}{\end{eqnarray}}
\newcommand{\mB}{m_B}
\newcommand{\mBst}{m_{B^*}}
\newcommand{\mBs}{m_{B_s}}
\newcommand{\mBsst}{m_{B_s^*}}

\def\eq#1{{Eq.~(\ref{#1})}}

\renewcommand{\logo}{\relax}

\title{The semileptonic $B_s$ and $\Lambda_b$ widths}

\author[a]{Marzia Bordone}
\author*[b]{Paolo Gambino}

\affiliation[a]{Theoretical Physics Department, CERN, 1211 Geneva 23, Switzerland}

\affiliation[b]{Dept. of Physics, University of Turin,\\
 via Giuria 1, 10125 Turin, Italy}

\emailAdd{gambino@to.infn.it}
\emailAdd{marzia.bordone@cern.ch}

\abstract{
 
We present estimates of the semileptonic widths of the $B_s$ and $\Lambda_b$ hadrons.  We employ the latest fit to the $B$ inclusive semileptonic data,  the heavy quark expansion and lattice QCD results. Our results suffer from large uncertainties that could be reduced with dedicated measurements and new lattice QCD calculations.}

\FullConference{%
  11th International Workshop on the CKM Unitarity Triangle (CKM2021)\\
  22-26 November 2021\\
  The University of Melbourne, Australia
}

\begin{document}
\begin{flushright}
{\small
CERN-TH-2022-050 
}
\end{flushright}
\maketitle

The determination of the Cabibbo-Kobayashi-Maskawa element $V_{cb}$  from inclusive semileptonic $B$ decays has been recently updated \cite{Bordone:2021oof} with the inclusion of new $O(\alpha_s^3)$ contributions to the width and to the relation between the $b$ quark mass in the $\overline{\rm MS}$ and in the kinetic scheme \cite{Fael:2020tow}, leading 
to 
\[ |V_{cb}|=42.16(51)\times 10^{-3} , \]
with a 30\% reduction in the total uncertainty. Indeed, the three-loop calculation brings the perturbative effects under better control. The new value for $|V_{cb}|$ implies a $\sim 4\sigma$ tension with the determination based on $B\to D^* \ell\nu$ data and lattice QCD \cite{FermilabLattice:2021cdg}.
The detailed results of the fit to inclusive semileptonic data can 
be found in Table I of Ref.~\cite{Bordone:2021oof}. In these 
proceedings we employ them to provide updated estimates of the 
total semileptonic widths of the $B_s$ and $\Lambda_b$, which can be useful in  semileptonic studies and in the determination of $f_s/f_d$ at LHCb.

\section{The semileptonic width}
As is well-known,  the semileptonic width can be written as a double expansion in $\alpha_s$ and $\Lambda_\mathrm{QCD}/m_b$,  \begin{equation}
\begin{aligned}
\label{eq1}
\Gamma_\mathrm{sl}(B_q) = \Gamma_0 f(\rho)\bigg[1&+a_1 a_s+ a_2a_s^2+a_3 a_s^3-\left(\frac{1}{2}-p_1 a_s\right)\frac{\mu_\pi^2(B_q)}{m_b^2}+(g_0+g_1 a_s)\frac{\mu_G^2(B_q)}{m_b^2}\\
&+(d_0 +d_1 a_s)\frac{\rho_D^3(B_q)}{m_b^3}-g_0 \frac{\rho_{LS}^3(B_q)}{m_b^3}+\dots\bigg]  \, ,
\end{aligned}
\end{equation}
where $a_s=\alpha_s(m_b)/\pi$ and  the subscript $q$ accounts for the down, up or strange spectator quark in the meson. The ellipses stand for higher order contributions not included in the reference fit of \cite{Bordone:2021oof}. We adopt the kinetic scheme for the $b$ quark mass and for the expectation values  
$\mu_\pi^2$, $\mu_G^2$, $\rho_D^3$ and $\rho_{LS}^3$, with cutoff $\mu=1\,\mathrm{GeV}$.  Unlike the perturbative corrections, the latter depend on the spectator quark and differ between $B\equiv B_{d,u}$ and $B_s$.
The charm quark mass is evaluated in the $\overline{\rm MS}$ scheme at $2\,\mathrm{GeV}$.
The current experimental data on the moments of the 
kinematic distributions in semileptonic $B$ decays allow us to extract the non-perturbative parameters for a light spectator, see Table I of  Ref.~\cite{Bordone:2021oof}. As  similar experimental data are not yet available for $B_s$ decays,  to estimate $\Gamma_\mathrm{sl}(B_s)$ one has to resort to  the 
Heavy Quark Expansion (HQE) of meson masses and to the heavy quark sum rules
\cite{Bigi:2011gf}. In the following, we update the analysis of Ref.~\cite{Bigi:2011gf} using the results of Ref.~\cite{Bordone:2021oof},
the latest PDG values for the meson masses \cite{ParticleDataGroup:2020ssz}, and recent lattice QCD data on meson masses obtained with both $d$ and $s$ spectators \cite{Gambino:2019vuo,Gambino:2017vkx}.

Before diving into the discussion, we report the contribution of the various non-perturbative parameters to the $B$ semileptonic width based on \cite{Bordone:2021oof}, relative to the lowest order $B$ width:
\begin{eqnarray}\label{eq:fit}
\frac{\delta_{\mu_\pi^2} \Gamma_\mathrm{sl}(B)}{\Gamma_\mathrm{sl}(B)}= -0.9(1) \%\,, && \frac{\delta_{\mu_G^2}\Gamma_\mathrm{sl}(B)}{\Gamma_\mathrm{sl}(B)} = -3.2(5) \%\,, \nonumber \\
\frac{\delta_{\rho_D^3}\Gamma_\mathrm{sl}(B)}{\Gamma_\mathrm{sl}(B)} = -3.2(5) \%\,, \ &&
\frac{\delta_{\rho_{LS}^3}\Gamma_\mathrm{sl}(B)}{\Gamma_\mathrm{sl}(B)} = -0.3(2) \%\,.
\end{eqnarray}
 Notice that the expectation values in (\ref{eq1}) refer to  physical $B$ mesons
and not to their infinite mass limit.

\section{SU(3) breaking effects in the HQE parameters}
We now follow closely the arguments of Ref.~\cite{Bigi:2011gf}.
For later convenience, we recall here the HQE for the heavy hadron masses
\be
M_H= m_b+ \bar\Lambda(H)+\frac{\tilde\mu^2_\pi(H)- c_G \,\tilde\mu^2_G(H)}{2m_b}+ O\Big(\frac1{m_b^2}\Big)\,,
\label{eq:masses}
\ee
where $\tilde\mu^2_{\pi,G}(H)$ are the expectation values in the infinite mass limit for the heavy quark. For $H=B$ they are related to the quantities entering Eq.~(1) by  $\mu^2_{\pi}(B)=\tilde\mu^2_{\pi}(B)- (\rho_{\pi\pi}^3+1/2 \rho^3_{\pi G})/m_b $ and  $\mu^2_{G}(B)=\tilde\mu^2_{G}(B)+ (\rho_{s}^3+\rho_A^3 +1/2 \rho^3_{\pi G})/m_b $, where the power suppressed contributions are given by expectation values of non-local operators that also appear in the $O(1/m_b^2)$ terms of Eq.~(\ref{eq:masses}).

The expectation value of $\tilde\mu_G^2$ can be extracted from the hyperfine splitting,
\begin{equation}
\frac{\tilde\mu^2_G(B_s)}{\tilde\mu^2_G(B)} \simeq \frac{\mBsst-\mBs}{\mBst-\mB} = 1.08 \pm 0.04\,,
\label{eq:muG_HQE}
\end{equation} 
up to power corrections. On the other hand, the HQE fits to meson masses computed on the lattice for different heavy quark masses and $q=d,s$ 

\cite{Gambino:2017vkx,Gambino:2019vuo} give 
\begin{equation}
\frac{\tilde\mu^2_G(B_s)}{\tilde\mu^2_G(B)} \simeq 1.20\pm 0.10\,.
\label{eq:muG_lattice}
\end{equation} 
 The fits of \cite{Gambino:2017vkx,Gambino:2019vuo} also find large power corrections to (\ref{eq:muG_HQE}), subject to substantial $SU(3)$ breaking. The numerical value given
in (\ref{eq:muG_HQE}) should therefore be considered with care. 
Moreover, what enters Eq.~(\ref{eq1}) are the expectation values in the physical mesons, 
while those appearing in the HQE of the meson masses refer to the infinite mass limit. 
We therefore  average the values in \eq{eq:muG_HQE} and \eq{eq:muG_lattice} and assign a large uncertainty that covers both values:
\begin{equation}
\frac{\mu^2_G(B_s)}{\mu^2_G(B)} \simeq 1.14\pm 0.10\,.
\label{eq:muG_lattice}
\end{equation} 
Recalling Eq.~(\ref{eq:fit}), we then expect 
\begin{equation}
\delta_{\mu_G^2}\frac{\Gamma_\mathrm{sl}(B_s)}{\Gamma_\mathrm{sl}(B)} =(-0.4\pm 0.3)\%\,.\label{eq:muG}
\end{equation}

Let us now consider the spin-averaged masses, defined by
\begin{equation}
\overline{m}_B = \frac{3 \mBst +\mB}{4}\,,
\end{equation}
and similarly for different spectators, in terms of which we get 
\begin{equation}
\begin{aligned}
\Delta m_B=\overline{m}_{B_s} -\overline{m}_B =&\,(89.9\pm1.2)\, \mathrm{MeV}\,,  & \Delta m_D=\overline{m}_{D_s} -\overline{m}_D =&\, (101.1\pm0.3)\, \mathrm{MeV}\,.
\end{aligned}
\label{eq:spinav_values}
\end{equation}
The spin-averaged masses admit  the  expansions 
\begin{equation}
\begin{aligned}
\overline{m}_{B} =&\, m_b + \bar\Lambda+\frac{\tilde\mu_\pi^2(B)}{2 m_b}+\mathcal{O}(1/m_b^2)\,, &
\overline{m}_{B_s} \simeq&\, m_b + \bar\Lambda_s+\frac{\tilde\mu_\pi^2(B_s)}{2 m_b}+\mathcal{O}(1/m_b^2)\,,
\end{aligned}
\label{eq:spinav_expansion}
\end{equation}
and similarly for the $D$ system, with the replacement $m_b\to m_c$. We can now use \eq{eq:spinav_values} and \eq{eq:spinav_expansion} to extract the values for $\bar\Lambda_s-\bar\Lambda$ and $\tilde\mu_\pi^2(B_s)-\tilde\mu_\pi^2 (B)$,
\begin{equation}
\bar\Lambda_s-\bar\Lambda = \frac{m_b \Delta m_B -m_c\Delta m_D}{m_b-m_c}\,, \qquad \tilde\mu_\pi^2(B_s)-\tilde\mu_\pi^2 (B) = \frac{2 m_b m_c}{m_b-m_c}(\Delta m_B-\Delta m_D)\,.
\end{equation}
Using the values of $m_{b,c}$ extracted in \cite{Bordone:2021oof}, we find
\begin{equation}
\bar\Lambda_s-\bar\Lambda \simeq 86\,\mathrm{MeV}\,, \qquad \tilde\mu_\pi^2(B_s)-\tilde\mu_\pi^2 (B) \simeq 0.032\, \mathrm{GeV}^2\,.\label{eq:12}
\end{equation}
This should be compared with the results reported in Ref.~\cite{Gambino:2019vuo}:
\begin{equation}
\bar\Lambda_s-\bar\Lambda \simeq (84\pm 20)\,\mathrm{MeV}\,, \qquad \tilde\mu_\pi^2(B_s)-\tilde\mu_\pi^2 (B) \simeq 0.11\pm 0.03\,\mathrm{GeV}^2\,.
\end{equation}
The discrepancy in $\tilde\mu_\pi^2(B_s)-\tilde\mu_\pi^2 (B)$ 
might again be explained by higher power corrections, which are neglected  in the derivation of \eq{eq:12}.
We take this and the different definition of the HQE parameters in (\ref{eq1}, \ref{eq:spinav_expansion}) 
into account  and use
\begin{equation}
\mu_\pi^2(B_s)-\mu_\pi^2 (B) = (0.1\pm 0.1)\,\mathrm{GeV}^2\,.
\label{eq:14}
\end{equation}
We can then estimate
\begin{equation}
\delta_{\mu_\pi^2}\frac{\Gamma_\mathrm{sl}(B_s)}{\Gamma_\mathrm{sl}(B)} =(-0.2\pm 0.1)\%\,.
\end{equation}

Let us now consider the contributions proportional to $\rho_D^3$.
Using the QCD equations of motion and the vacuum saturation approximation one expects 
\begin{equation}
\frac{\rho_D^3(B_s)}{\rho_D^3(B)} \simeq \frac{f_{B_s}^2\mBs}{f_B^2 \mB} = 1.48\pm 0.06 \,,
\label{eq:rhoD1}
\end{equation}
where the value $f_{B_s}/f_B = 1.206\pm0.023$ has been used \cite{
FlavourLatticeAveragingGroup:2019iem}.  We can also use the heavy quark sum rules to estimate the same ratio
\begin{equation}
\frac{\rho_D^3(B_s)}{\rho_D^3(B)} \simeq \left(\frac{\mu^2_\pi (B_s)}{\mu^2_\pi (B)}\right)^2 \frac{\bar\Lambda}{\bar\Lambda_s} = 1.30 \pm 0.10\,,
\label{eq:rhoD_SR}
\end{equation}
where we employed (\ref{eq:14}) and the results of \cite{Bordone:2021oof}. Averaging the values in \eq{eq:rhoD1} and \eq{eq:rhoD_SR} we arrive at
\begin{equation}
\frac{\rho_D^3(B_s)}{\rho_D^3(B)} \simeq 1.39 \pm 0.15\,.
\label{eq:rhoD_final}
\end{equation}
The fits of Refs.~\cite{Gambino:2017vkx,Gambino:2019vuo} do not extract $\rho_D^3$, but  a linear combination of different terms:
\begin{equation}
\frac{\rho_D^3(B_s)+\rho^3_{\pi\pi}(B_s) -\rho_S^3(B_s) }{\rho_D^3(B)+\rho^3_{\pi\pi}(B) -\rho_S^3(B)} = 1.33 \pm 0.25\,,
\end{equation}
which confirms a sizeable increase of the $1/m_b^3$ terms. In summary,
the impact of \eq{eq:rhoD_final} on the semileptonic width is quantifiable as 
\begin{equation}
\delta_{\rho_D^3}\frac{\Gamma_\mathrm{sl}(B_s)}{\Gamma_\mathrm{sl}(B)} =(-1.2\pm 0.6)\%\,. \label{eq:rhoD}
\end{equation}
Since the impact of $\rho_{LS}^3$ on the rate is rather small, we will not discuss its $SU(3)$ breaking.
Concerning  higher power corrections, we have indications that $O(1/m_Q^4)$ and $O(1/m_Q^5)$ are suppressed with respect to lower power corrections \cite{Gambino:2016jkc}, but  
they might be more sensitive to $SU(3)$ breaking. We therefore assign an additional 0.5\% uncertainty to the ratio of the semileptonic widths, which together with 
 Eqs.~(\ref{eq:muG},\ref{eq:14},\ref{eq:rhoD}) leads to our final estimate
\begin{equation}
\frac{\Gamma_\mathrm{sl}(B_s)}{\Gamma_\mathrm{sl}(B_d)}-1 = - (1.8 \pm 0.8)\% \,. \label{eq:finalBs}
\end{equation}

\section{The $\Lambda_b$ case}
The $\Lambda_b$ case can be dealt with in a similar fashion, 
keeping in mind that in the $\Lambda_b$ baryon  $\mu_G^2$ and $\rho_{LS}^3$ vanish at the leading order in $1/m_b$.
Subtracting these terms will give the largest difference in the semileptonic width. From the fit in Ref.~\cite{Bordone:2021oof} we get
\begin{equation}
\frac{ \delta_{\mu_G^2} \Gamma_\mathrm{sl}(B)+ \delta_{\rho_{LS}^3} \Gamma_\mathrm{sl}(B)} {\Gamma_\mathrm{sl}(B)} =
-(3.5\pm 0.6)\%
\,.
\end{equation}
For the other contributions we basically confirm the estimates of Ref.~\cite{Bigi:2011gf} (see also
\cite{Colangelo:2020vhu}) and find 
\begin{equation}
\delta_{\mu_\pi^2}\ \frac{\Gamma_\mathrm{sl}(\Lambda_b)}{\Gamma_\mathrm{sl}(B)}  = (-0.2 \pm 0.2)\%\,, \quad \delta_{\rho_D^3}\frac{\Gamma_\mathrm{sl}(\Lambda_b)}{\Gamma_\mathrm{sl}(B_d)}  = (0.8 \pm 1.3)\%\,.
\end{equation}
Putting together the above results and including  an additional 0.7\% uncertainty due to subleading effects of the chromomagnetic operator and to higher power corrections, 
we find
\begin{equation}
\frac{\Gamma_\mathrm{sl}(\Lambda_b)}{\Gamma_\mathrm{sl}(B_d)}-1 = (4.1 \pm 1.6)\% \,,
\end{equation}
where the large uncertainty 
is dominated by the value of $\rho_D^3(\Lambda_b)$.

\section{Summary}
We have estimated the $B_s$ and $\Lambda_b$ semileptonic widths, updating Ref.\cite{Bigi:2011gf} in various ways.
These semileptonic widths are an important input in a few LHCb analyses.
The leading uncertainty in our estimates is related to the Darwin term $\rho_D^3$ and only a better knowledge of this parameter in the $B_s$ and $\Lambda_b$ can improve them. In particular, a  measurement of the moments of kinematic distributions
such as the hadronic mass distribution in $B_s$ semileptonic decays at LHCb could  improve the current theoretical prediction of $\Gamma_\mathrm{sl}(B_s)$.

The inclusive semileptonic width can now also be computed on the lattice 
\cite{Gambino:2020crt,Gambino:2022dvu}, but we do not have yet
results for physical values of $m_b$ and a complete assessment of the lattice systematic uncertainties is still missing. In the future, however, the ratio $\Gamma_\mathrm{sl}(B_s)/\Gamma_\mathrm{sl}(B)$ might be determined with an accuracy similar or better than in (\ref{eq:finalBs})
as  many systematic uncertainties will cancel out between numerator and denominator. For what concerns the $\Lambda_b$ semileptonic width, an improvement could come from a study
similar to that of Ref.~\cite{Gambino:2017vkx} for heavy baryons.

\vspace{4mm}

This work is supported in part by the Italian Ministry of Research (MIUR) under grant PRIN 20172LNEEZ.

\end{document}